\documentclass[useAMS,usenatbib]{mn2e}
\usepackage[dvips]{graphicx}
\usepackage{subfigure}

\title[Evolution of the Radio Loud Galaxy Population]{Evolution of the Radio
Loud Galaxy Population}
\author[Donoso, Best and Kauffmann]{E. Donoso$^{1}$\thanks{E-mail:edonoso@mpa-garching.mpg.de} P. N. Best$^{2}$ and G. Kauffmann$^{1}$\\
$^{1}$Max-Planck-Institut f\"{u}r Astrophysik, Karl-Schwarzschild-Str. 1, Postfach 1317, D-85741 Garching, Germany\\
$^{2}$SUPA, Institute for Astronomy, Royal Observatory Edinburgh, Blackford Hill, Edinburgh EH9 3HJ, UK}

\begin{document}
\date{Accepted ???? ??? ??. Received ???? ??? ??}
\pagerange{\pageref{firstpage}--\pageref{lastpage}} \pubyear{2008}
\maketitle
\label{firstpage}

\begin{abstract}
A catalogue of 14453 radio--loud AGN with 1.4 GHz fluxes above 3.5 mJy 
in the redshift range $0.4<z<0.8$, has been constructed from the
cross--correlation of the NVSS and FIRST radio surveys with the MegaZ--LRG
catalogue of luminous red galaxies derived from Sloan Digital Sky Survey imaging
data. The NVSS provides accurate flux measurements for extended sources, 
while the angular resolution of FIRST allows the host galaxy to be identified
accurately. New techniques were developed for extending the cross--correlation
algorithm to FIRST detections that are below the nominal 1 mJy S/N limit of the
catalogued sources. The matching criteria were tested and refined using
Monte--Carlo simulations, leading to an  estimated reliability of $\sim$98.3\%
and completeness level (for LRGS) of about 95\% for our new catalogue.

We present a new determination of the luminosity function of radio AGN at $z \sim 0.55$
and compare this to the luminosity function of nearby ($z \sim 0.1$) radio
sources from the SDSS main survey. The comoving number density of radio AGN with
luminosities less than $10^{25}$ W Hz$^{-1}$  increases by a factor  $\sim 1.5$
between $z =0.1$ and $z=0.55$. At higher lumiosities, this factor increases
sharply, reaching values of more than 10 at radio luminosities larger than
$10^{26}$ W Hz$^{-1}$. We then study how the relation between radio AGN and their
host galaxies evolves with redshift. Our main conclusion is that the fraction of
radio--loud AGN increases towards higher redshift in all massive galaxies, but
the evolution is particularly strong for the lower mass galaxies in our sample.
These trends may be understood if there are two classes of radio galaxies
(likely associated with the ``radio"  and ``quasar mode" dichotomy) that
have different fuelling/triggering mechanisms and hence evolve in different ways.
\end{abstract}

\begin{keywords}
galaxies: evolution -- galaxies: active -- radio continuum: galaxies.
\end{keywords}

\section{Introduction}
Ever since the first studies of \citet{longair}, it has been widely accepted that 
evolution is required to explain the radio source counts. Strong cosmological
evolution of the high luminosity radio source population has been established out
to at least $z=2-3$ (e.g. \citealt{dunpeack}), with the space--densities of
the most powerful radio sources increasing by a factor $\sim10^3$ relative to
their densities at $z\sim0$. However, the evolution of the low--luminosity radio
galaxy population remained more uncertain.
\citet{clewley} found that the low luminosity population of radio sources with
$L_{\rm 325 MHz} < 10^{25}$ W Hz$^{-1}$sr$^{-1}$ exhibits no evolution out to  
$z\sim0.8$. In contrast, \citet{brown} found significant
cosmic evolution out to $z \sim 0.55$, which they modelled using a pure
luminosity evolution model as $P(z) = P(0)(1+z)^{k_L}$, with $3<{k_L}<5$. More
recently, similar results were obtained by \citet{sadler07}, who analyzed radio
galaxies in the `2dF--SDSS Luminous Red Galaxy and Quasar' (2SLAQ) survey
(\citealt{cannon}). They found that the number density of radio sources below
$\rm P_{\rm 1.4 GHz}=10^{26}$ W Hz$^{-1}$ grows by a factor of $\sim 2$ between
$z=0$ and $z=0.55$. These results confirm that the cosmic evolution of
low power radio sources is considerably weaker than that of their higher radio
power counterparts. However, because of the relatively small sizes of samples, none of
these studies was able to accurately constrain the luminosity dependence of the
evolution.

Over the same period, our physical understanding of radio galaxies has developed
steadily. \citet{fanaroff} showed that radio galaxies exhibit a dichotomy in
radio morphology. Low radio luminosity sources are usually `edge--darkened', with
radio jets that quickly decelerate and flare as they advance through the
interstellar medium (Fanaroff--Riley class I sources, or FRIs). In contrast, most
higher luminosity sources have jets that remain relativistic over kpc and up to Mpc
scales, ending in bright hotspots (`edge--brightened' sources, or FRIIs). The
dividing line between FRIs and FRIIs occurs around a monochromatic power of
log$_{10}$(P$_{\rm 1.4 GHz}$)$\sim25$ W Hz$^{-1}$, although \citet{ledlowowen}
showed that it occurs at higher radio luminosity in more optically--luminous
galaxies. This division is close to the break in the radio luminosity function,
and has led to the hypothesis these two populations of radio galaxies correspond
to the different populations of cosmologically--evolving sources, FRIIs being
strongly evolving and FRIs much less so (e.g. \citealt{jacksonwall}).

A more important division in the radio source population, however, appears
to relate to the radiative properties of the AGN. Powerful FRIIs and a
small subset of the more powerful FRIs (e.g. see \citealt{heywood}) show
strong emission lines and either a visible or hidden quasar--like nucleus
(\citealt{barthel}), indicative of radiatively efficient accretion. In contrast,
most FRIs and a sub--population of weak--emission--lined low luminosity
FRIIs show no evidence for a powerfully radiating nucleus (see
\citealt{hardcastle} and references therein) suggesting radiatively
inefficient accretion. This implies that a fundamental difference exists
between high and low power radio sources that is distinct from the radio
morpological dichotomy. Instead, it is likely to be related to different
origins or mechanisms of gas fuelling.

Understanding this low luminosity radio source population, and its cosmic
evolution, is not only important for understanding the physical origin of the
radio activity, but also for understanding galaxy evolution. There is growing
evidence that low--luminosity radio--loud AGN play a critical role in regulating
the masses and star formation rates of galaxies. Semi--analytic models of galaxy
formation  (for example see \citealt{whitefrenk} and \citealt{cole}) successfully
reproduce many of the  the observed properties of galaxies. However, these models
over--predict the abundance of galaxies at the bright end of the luminosity
function. This problem arises because many massive galaxies are predicted to sit
at the centre of hot hydrostatic haloes, in which cooling flows are expected to
develop. If the gas cools and forms stars, central group and cluster galaxies are
too massive by the present day and have much bluer colours than observed.
\citet{tabor} first suggested that radio galaxies could in principle regulate
these cooling flows, preventing significant accretion of gas and limiting the mass
of galaxies. This concept of AGN feedback has now been successfully incorporated
into the semi--analytic models of \citet{bower} and \citet{croton}.

Observational support for a scenario in which radio AGN play an important role
in regulating the growth of massive galaxies has slowly been accumulating. Both
luminous FRII and low power FRI sources can in principle inject a significant
amount of energy into the surrounding gas, as their radio lobes expand and
interact with the surrounding medium ($\sim10^{42}$ ergs$^{-1}$ for the jets
of a borderline FRI/FRII object, see \citealt{bicknell}). This has been directly
observed in clusters of galaxies, where the expanding radio lobes create buoyant
bubbles and drill cavities in the hot intracluster medium (\citealt{bohrin};
\citealt{mcnam}; \citealt{fabian}). In nearby systems, which can be studied in
detail, the energy supplied by the radio source is well--matched to that
required to balance the cooling (e.g. \citealt{fabian}).

With the availability of large redshifts surveys like the 2--degree--Field Galaxy
Redshift Survey (2dFGRS, \citealt{colless}) and the Sloan Digital Sky Survey
(SDSS, \citealt{york}) it has become possible to obtain the sky and redshift
coverage needed to define sufficiently large samples of nearby radio sources to
study population statistics and global energetics. \citet{best05a} constructed a
sample of 2215 radio--loud AGN brighter than 5 mJy with redshifts $0.03<z<0.3$
using the SDSS Data Release 2 (DR2) in combination with the National Radio
Astronomy Observatory Very Large Array Sky Survey (NVSS, \citealt{condon}) and
the Faint Images of the Radio Sky at Twenty centimeters (FIRST, \citealt{becker}).
\citet{best05b}  found that the fraction of radio--loud AGN, $f_{RL}$, is a
strong function of black hole and stellar mass, with dependencies
$f_{RL}\propto M_{*}^{2.5}$ and $f_{RL}\propto M_{BH}^{1.6}$. Twenty--five percent
of the very most massive galaxies host radio sources more powerful than
$10^{23}$ W Hz$^{-1}$. By combining these results with estimates of the energetic
output of these sources, \citet{best06} showed that the time--averaged
mechanical energy output of radio sources in elliptical galaxies is
comparable to the radiative cooling losses from the hot halo gas, for
ellipticals of all masses. Radio sources can therefore provide sufficient
energy to suppress cooling flows and regulate the growth of massive
galaxies in the nearby Universe.

The role of powerful (radiatively--efficient) FRII objects could not be
addressed by Best et~al. due to the scarcity of such systems in the low redshift
catalogue. Another  open question is how the heating--cooling balance in
massive galaxies evolves with redshift. In order to address these issues, we
require a large radio galaxy sample at higher redshifts. In this paper we extend
the work of Best et~al. by constructing a comparable catalogue of radio--loud
AGN  over the redshift range $0.4 < z < 0.8$. The catalogue contains a large
number of luminous radio sources. By comparing with a sample of local radio sources
selected  from SDSS DR4, we aim to  study not only of the evolution of the radio
galaxies, but also the evolution of the duty cycle of radio AGN activity in
galaxies as a function of their stellar mass.

This paper is organized as follows. In Section 2 we will describe the surveys
and samples used in this work, as well as the matching procedure. In Section 3
we will present results on the evolution of the radio luminosity function,
radio--loud AGN fractions and the bivariate radio luminosity--stellar mass
function, and the stellar mass and colour distributions of radio AGN host
galaxies. Finally, we will summarize our results in Section 4 and discuss the
implications of our work.

\section{The Radio--Optical Galaxy Samples}
\subsection{The MegaZ--LRG Galaxy Catalogue}
The Sloan Digital Sky Survey (\citealt{york}; \citealt{stoughton}) is a
five--band photometric and spectroscopic survey that has mapped almost a
quarter of the whole sky, providing precise photometry for more than 200
million objects and accurate redshifts for about a million galaxies and
quasars. Imaging data in the {\it ugriz} bands (\citealt{fukugita}) has been
obtained with a large format CCD camera (\citealt{gunn}) mounted on a
dedicated telescope located at Apache Point Observatory in New Mexico.
Astrometric coordinates
are accurate  to $\sim0.1$ arcsec per coordinate. Throughout this work,
unless otherwise stated, all magnitudes are model magnitudes, derived from the
best--fit de Vaucouleurs or exponential profile 
in the {\it r} band. The amplitude is then scaled to fit
measurements in other filters. This provides the best estimate of galaxy colours
since the same aperture is used in all passbands. A correction for foreground Galactic
extinction is applied to all the magnitudes following \citet{schlegel}.

The MegaZ--LRG (\citealt{collist}) is a photometric redshift catalogue
based on imaging data from  the fourth Data Release (DR4) of the SDSS. It 
consists of  $\sim$1.2 million Luminous Red Galaxies (LRGs) with limiting
magnitude  ${\it i} < 20$ over the redshift range $0.4 < z < 0.8$. MegaZ adopts
various colour and magnitude cuts to isolate red galazies at $0.4 < z < 0.8$. The
cuts are very similar to those adopted by the `2dF--SDSS LRG and Quasar' project
(2SLAQ, \citealt{cannon}). The specific colour cuts are:

\begin{equation}
0.5 < g-r < 3
\end{equation}

\begin{equation}
r-i < 2
\end{equation}

\begin{equation}
c_{par} \equiv 0.7(g-r) + 1.2(r-i-0.18) > 1.6
\end{equation}

\begin{equation}
d_{perp} \equiv (r-i) - (g-r)/8 > 0.5
\end{equation}

Essentially, $c_{par}$ isolates early--type galaxies (LRGs), while $d_{perp}$
selects galaxies with redshifts $z > 0.4$. Accurate photometric redshifts are available
for the entire LRG sample. These are derived using a  neural network photometric redshift
estimator (ANNz, \citealt{collistlah}). The neural net was trained using a sample
of $\sim13000$ LRGs with  spectroscopic redshifts from 2SLAQ. The r.m.s. average
photometric redshift error for all the galaxies in the sample is $\sigma_{rms}=0.049$.

\subsection{The NVSS and FIRST Radio Catalogues}
The NRAO VLA Sky Survey (NVSS; \citealt{condon}) is a radio continuum
survey at 1.4 GHz, covering the whole sky at declinations $\delta_{J2000} >
-40^{\circ}$. The survey provides total intensity and polarization measurements down to a
limiting point source flux density of $\sim2.5$ mJy. The rms noise and
confusion limit in the total intensity maps is $\sigma_{RMS}\approx0.45$
mJy beam$^{-1}$, except near very strong sources. The survey was carried
out using the VLA in compact D and DnC configurations, which provides an
angular resolution of around 45 arcsec. A catalogue of $\sim1.8$ million
point sources was derived by fitting elliptical gaussians to the discrete
peaks in the images. The catalogue lists peak flux densities and best--fit  sizes for all
sources with fluxes $S_{peak} \geq 2$ mJy beam$^{-1}$. The completeness of the
NVSS survey is around 50\% at $\sim2.5$ mJy,  but rises sharply to more than 99\%
at 3.5 mJy.

The VLA Faint Images of the Radio Sky at Twenty Centimeters (FIRST;
\citealt{becker}) is a survey that covers 10000 deg$^2$ of the North
Galactic Cap, using the VLA in its B configuration at 1.4 GHz. This
delivers a resolution of $\sim$5 arcsec and a typical noise of
$\sigma_{RMS}\approx0.13$ mJy, except in the vicinity of bright sources
($>100$ mJy) where sidelobes can lead to an increased noise level. The sky
coverage overlaps most of the SDSS survey in the Northern
Galactic Cap. As with the NVSS, a source catalogue including peak and integrated flux
densities and sizes was derived from fitting a two--dimensional Gaussian to
each source generated from the co--added images. Astrometric accuracy in
this source catalogue is better than 1 arcsec for the faintest sources
detected at the nominal threshold of 1 mJy.

\subsection{Identification of radio--loud AGN}
The NVSS and FIRST surveys are highly complementary in many respects. NVSS has sufficient
surface brightness sensitivity to provide accurate flux measurements of extended
radio sources with lobes and jets. On the other hand, the superior angular
resolution of FIRST is necessary to identify the central core component of the
radio sources and associate it with the host galaxy detected in optical surveys.

\citet{best05a} developed an automated hybrid cross--matching method to
identify radio galaxies in the SDSS DR2. The method made use of both NVSS
and FIRST, thereby exploiting the
advantages of each survey and avoiding the errors that arise by
using them separately. A collapsing algorithm  was used to identify
multiple--component FIRST and NVSS sources. 
The final catalogue consisted of 2712 radio galaxies 
brighter than 5 mJy (2215 radio--loud AGN
and 447 star forming galaxies). The reliability and completeness
of the catalogue was estimated to be  98.9\% and 95\%, respectively.

This paper presents an extension of the \citet{best05a} method by combining NVSS and
FIRST with the high redshift MegaZ--LRG catalogue. We largely follow the procedure 
described in \citet{best05a}, but a few small modifications were implemented
in order to accommodate the characteristics of the new sample. The changes were
tested and progressively refined using a set of 10 Monte--Carlo simulations of
a 720 deg$^2$ patch of the survey area. The numbers quoted in the description
below pertain to this small test area.

\begin{enumerate}
\item \citet{best05a} considered radio sources above a limit of 5 mJy.
The major reason for this choice  was to reduce any loss of sensitivity to
multi--component NVSS sources. The MegaZ radio sources are more distant and
their angular sizes are thus smaller, so there are fewer multi--component
sources (see Section 2.5). In order to probe fainter radio galaxies out to higher
redshifts, we adopted a flux limit of 3.5 mJy. This limit corresponds to
approximately 7 times the noise level in the NVSS maps.

\item Figure 1 shows the distribution of positional offsets between the SDSS LRG and
the nearest NVSS match, for galaxies not classified as multi--component NVSS
sources. There is a clear excess of galaxies compared to
random matches for separations less  $\sim$10-15 arcsec. The excess
remains significant out to $\sim$70 arcsec. A significant part of the effect
at larger angular separations is because galaxies are intrinsically clustered
(see discussion in \citet{best05a} and also Section 2.5). However, some part
may consist of true associations between LRGs and 
extended NVSS sources. The subsequent FIRST cross--matching helps us pinpoint
such cases. \citet{best05a} carried out this cross--matching for all
SDSS galaxies with a single--component NVSS match within 30 arcsec. In this work,
this limit is increased to 60 arcsec.  This identifies an extra 37 (2.8\%) genuine
radio galaxies  at a cost of 0.8 false matches, i.e. a decrease in reliability of only 
$\sim$0.01\%.
\item Because the NVSS radio sources reach a fainter limit than those studied
by \citet{best05a}, we find more objects ($\sim$8\%) with no FIRST match. These
might either be extended sources missed by the low surface brightness sensitivity
of FIRST, or because the radio source has faded between observations. A careful
inspection of FIRST images (along with NVSS and SDSS cutouts) around
the positions of these objects revealed that in many cases there were 
bright spots near or slightly below the survey flux limit, that were clearly
associated with a MegaZ optical galaxy. A method to use this information to
increase the number of matches was developed and tested. The details are
explained in the following subsection.
\end{enumerate}

\begin{figure}
\includegraphics[width=84mm]{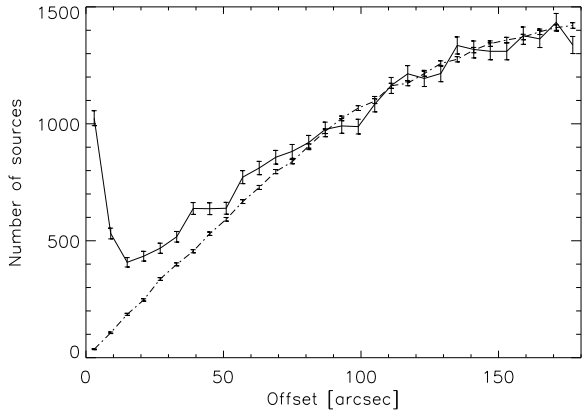}
\caption{Distribution of offsets between the MegaZ--LRG galaxies and the
closest NVSS source (solid line) and the corresponding offset distribution
derived from the random samples (dot--dashed line). Most true matches clearly
stand out at offsets less than 15 arcsec, but the excess remains
significant out to separations of  $\sim$70 arcsec.} 
\end{figure}

\subsection{Improved matching of sources without FIRST counterparts}
The high resolution of the FIRST survey is essential to allow reliable
cross--matching with dense optical samples. The NVSS sources without FIRST
counterparts are the largest source of incompleteness in the final sample (see
\citet{best05a}). However, blindly accepting all LRG candidates out to a
separation of 20 arcsec from the NVSS source introduces an average of 42.2 false
detections in our random samples, reducing the reliability of the catalogue to 79\%
within our test area (see Figure 2, where the distribution of NVSS--MegaZ
offsets are plotted for real and random data).

We turned to the analysis of the FIRST radio maps around the candidate positions,
in order to retain as many of these sources as possible without introducing
significant contamination. Figure 3 shows FIRST cutouts of typical objects with
no catalogued FIRST source, but with a nearby NVSS--MegaZ association. In
many cases one or more bright ``spots" matching the NVSS source and MegaZ
galaxy can be seen, while in some cases two well--defined symmetric radio
lobes centered around the optical galaxy are evident.

\begin{figure}
\includegraphics[width=84mm]{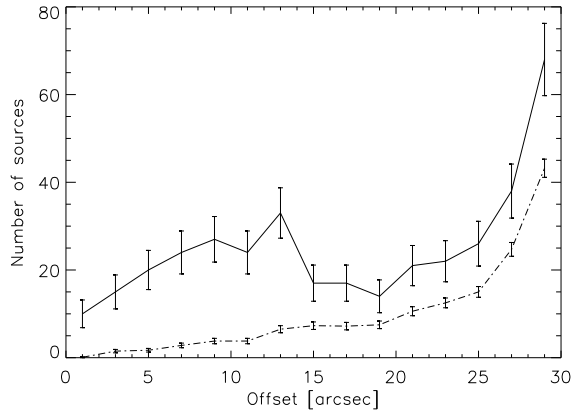}
\caption{Distribution of offsets between MegaZ--LRG galaxies and NVSS candidates
for sources that have no match  in the FIRST catalogue. Results
are shown for real (solid line)
and random (dot--dashed line) samples. Matches with separations larger
than  $\sim$15 arcsec are mostly the result of galaxy clustering.} 
\end{figure}

We took advantage of this additional information in deciding whether
candidates should be accepted  as true FIRST matches. 
For galaxies within 10 arcsec of the NVSS source we accepted all matches,
following \citet{best05a}. In the test area, this provided 96 associations of
which 9.9 are expected to be false (i.e. 10.3\% contamination). For galaxies
with offsets between 10 and 15 arcsec, we accepted matches according to
the following procedure:

\begin{enumerate}
\item
The FIRST image was segmented using an initial threshold of 
$3 \sigma_{rms}(\alpha,\delta)$, where
$\sigma_{rms}(\alpha,\delta)$ is the {\it local} rms noise in the map.
This provided a number of seed spots whose
centroids were refined using  the intensity--weighted moments in both
directions.

\item
A brightness profile was
derived for each spot and both the half light radius and the flux
inside this radius were determined.

\item
Any bright spot with a mean flux density greater than 0.4 mJy and within 5
arcsec of the optical counterpart was considered to be a valid match.
These sources were accepted unless there were more than 4 bright spots
inside a circle of 30 arcsec, in which case the match was rejected because
it is then impossible to associate the source with a single MegaZ galaxy.
Visual inspection of SDSS images confirmed this.
\end{enumerate}

To test this 'spot matching' technique we repeated the same procedure
for matches without FIRST components in the random
samples. This constitutes a robust check, since we are in effect looking at 
random locations in the sky, so we can assess how probable it is to find a
spot pattern by chance. In the sky test area, our procedure identified 13
additional FIRST for  sources with NVSS--MegaZ offsets between
10 and 15 arcsec, at the cost of
introducing only 2.4 false detections. This increases  the
overall completeness of the sample, for little cost in reliability. 
Our tests indicate that it is not possible to  extract more 
matches beyond a $\sim$15 arcsec offset with reasonable confidence.
\begin{figure*}
     \centering
     \subfigure[]{
          \includegraphics[width=.30\textwidth]{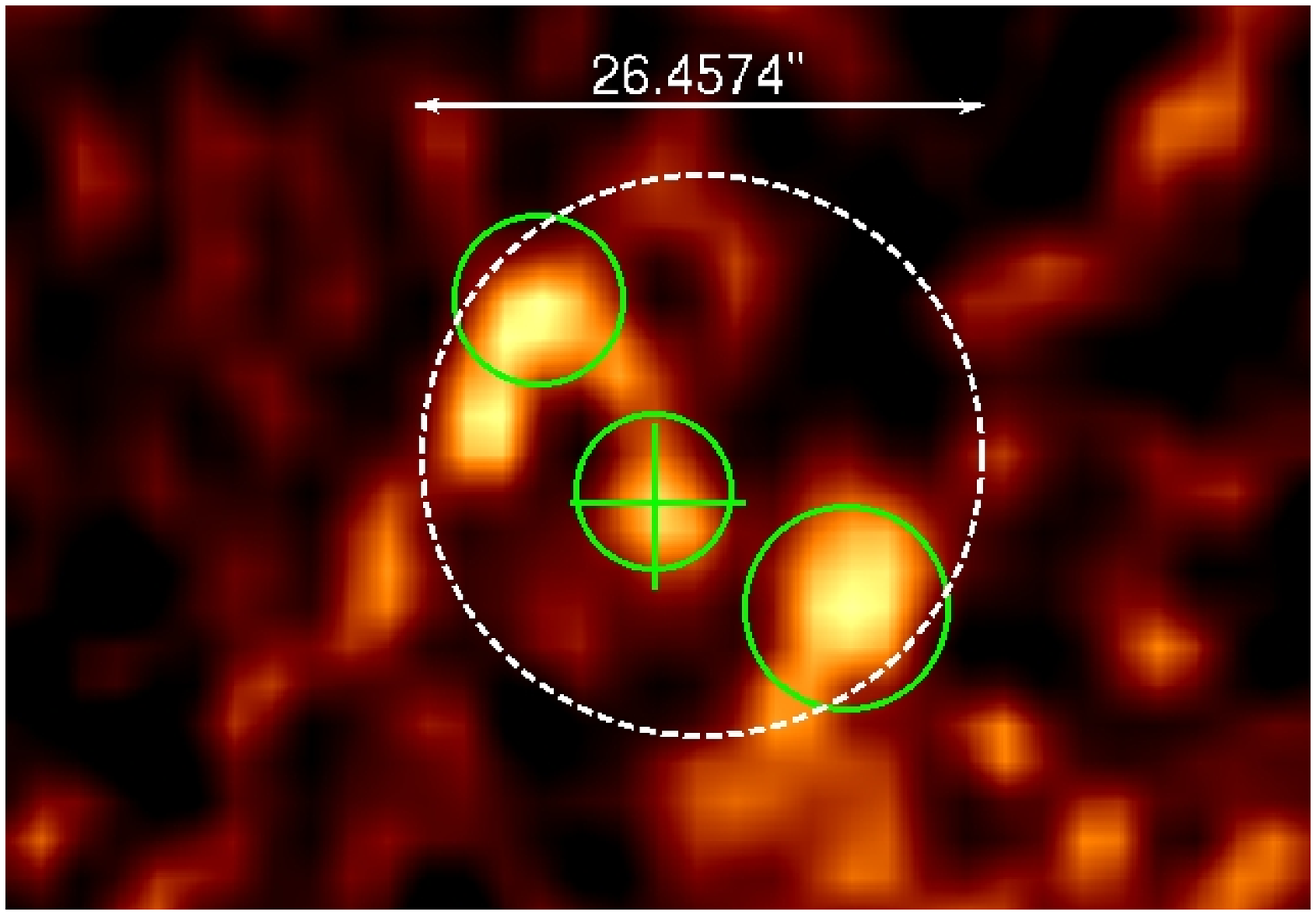}}
     \subfigure[]{
          \includegraphics[width=.30\textwidth]{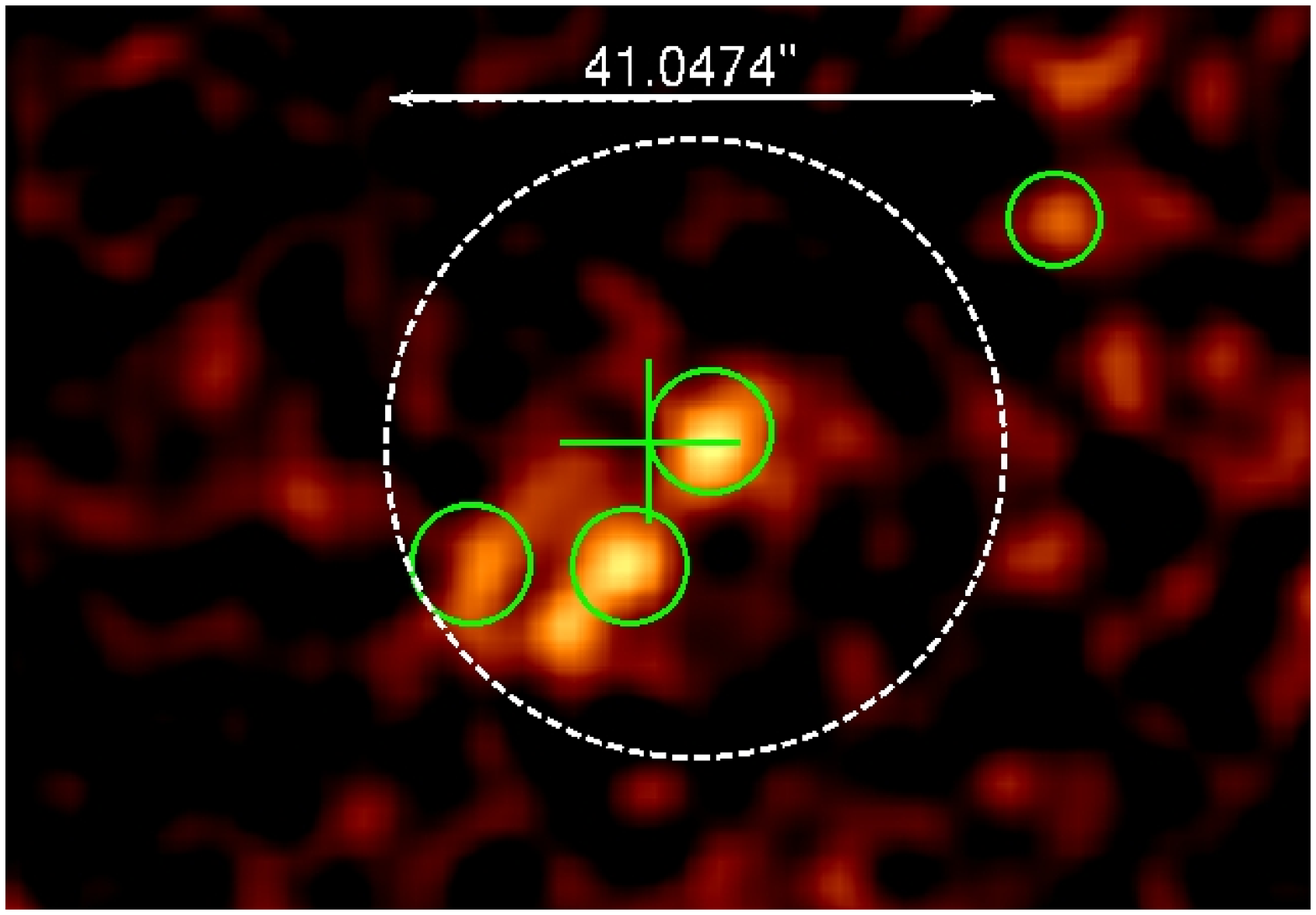}}
     \subfigure[]{
          \includegraphics[width=.30\textwidth]{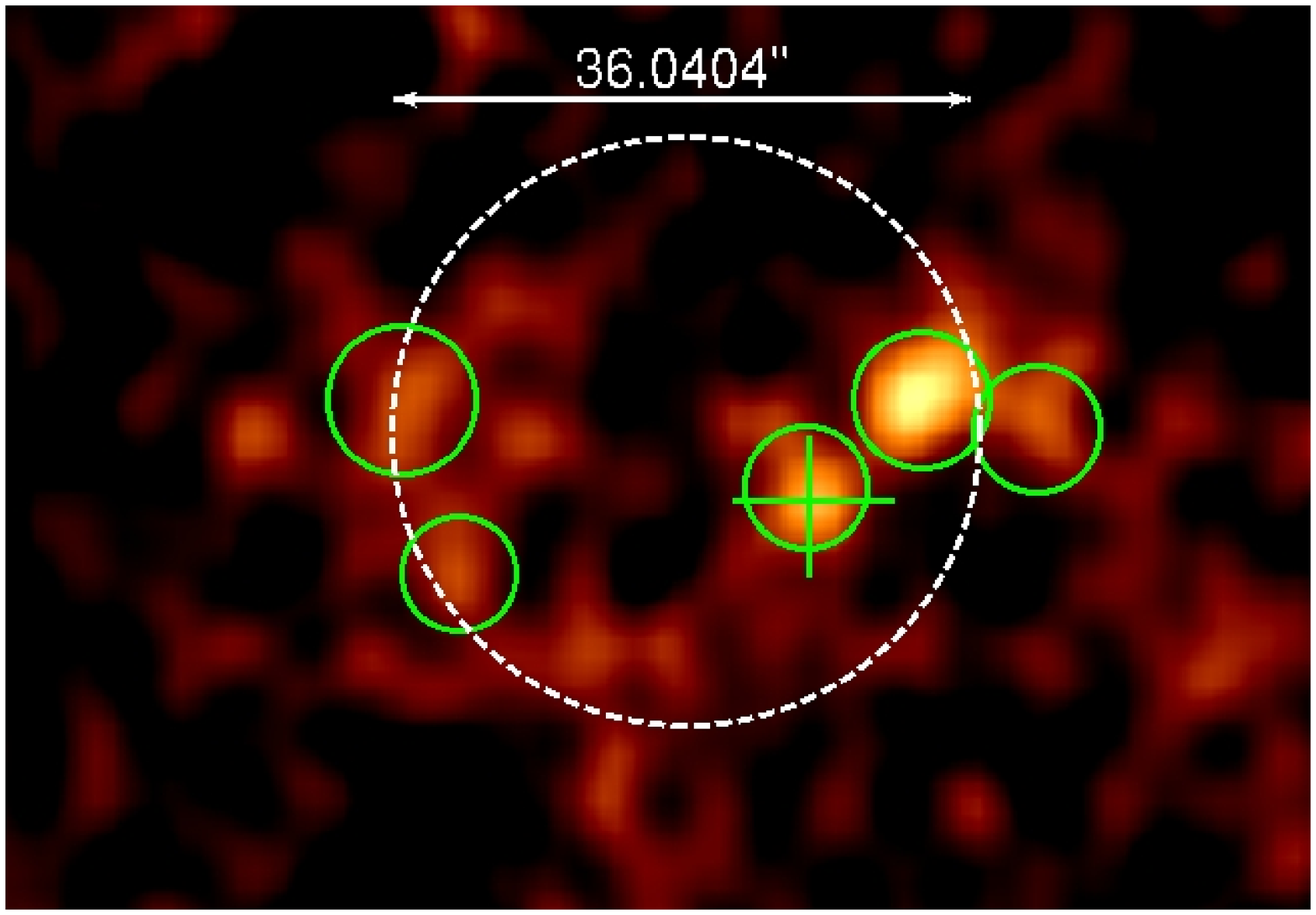}}\\
     \subfigure[]{
          \includegraphics[width=.30\textwidth]{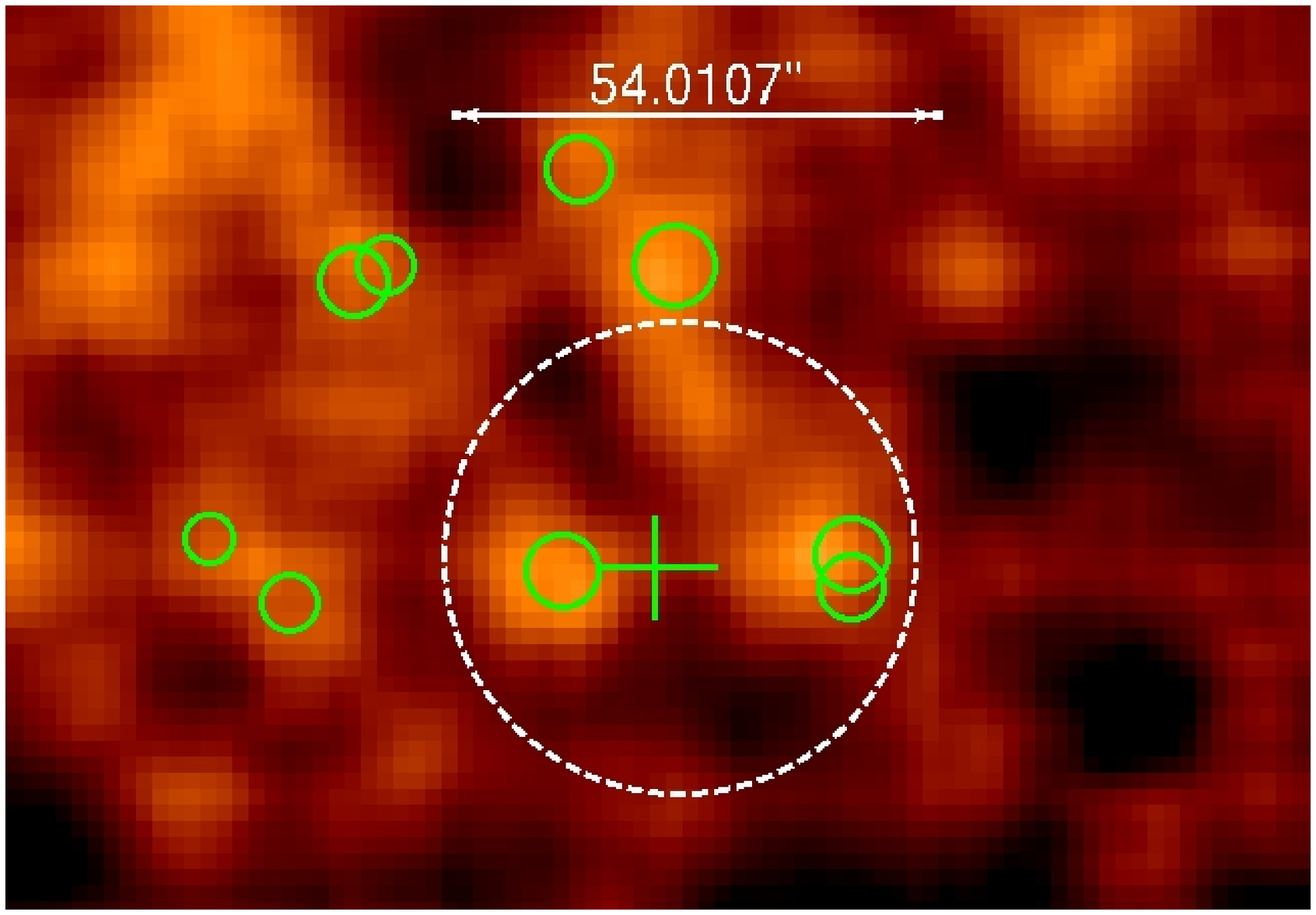}}
     \subfigure[]{
          \includegraphics[width=.30\textwidth]{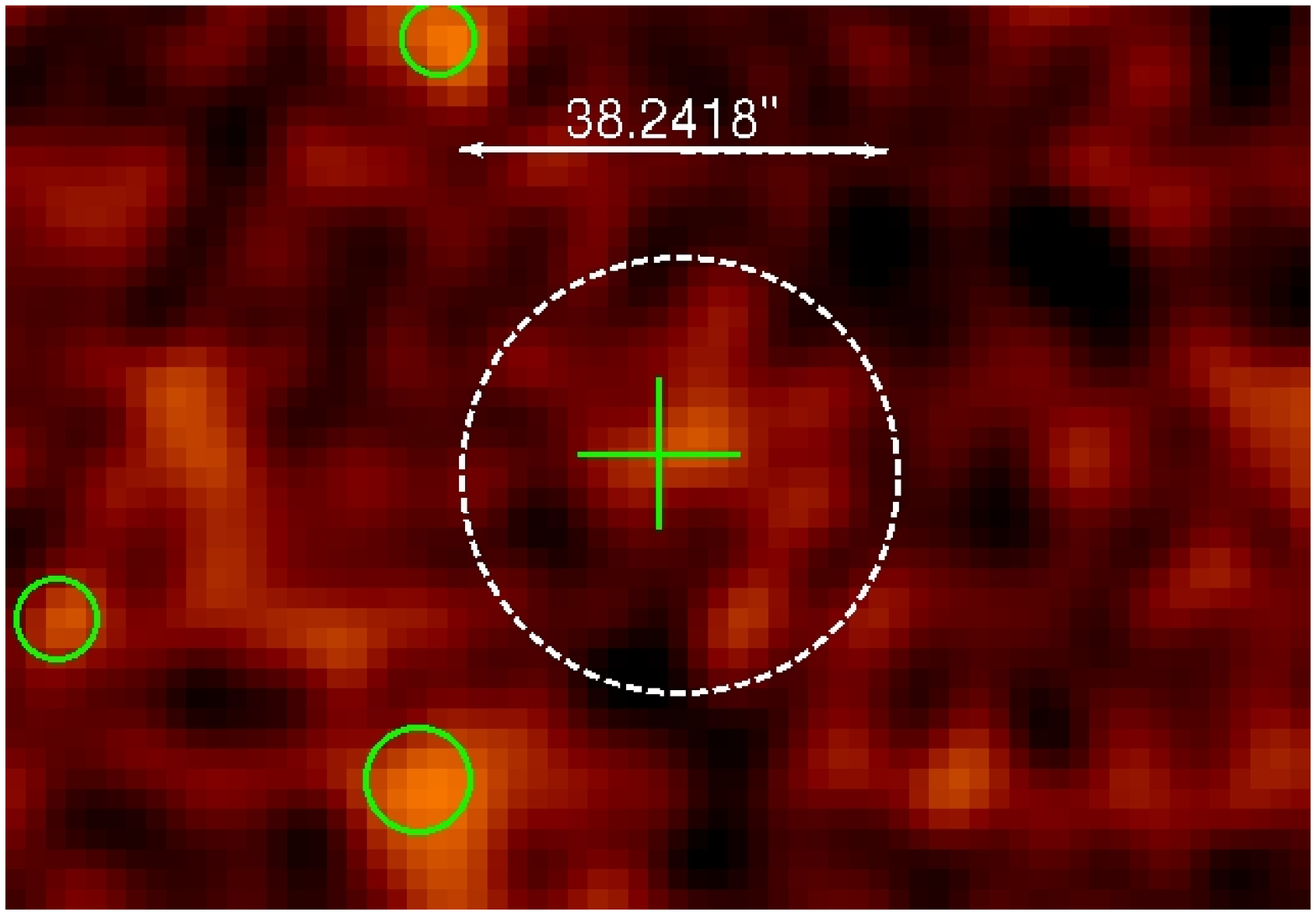}}
     \subfigure[]{
          \includegraphics[width=.30\textwidth]{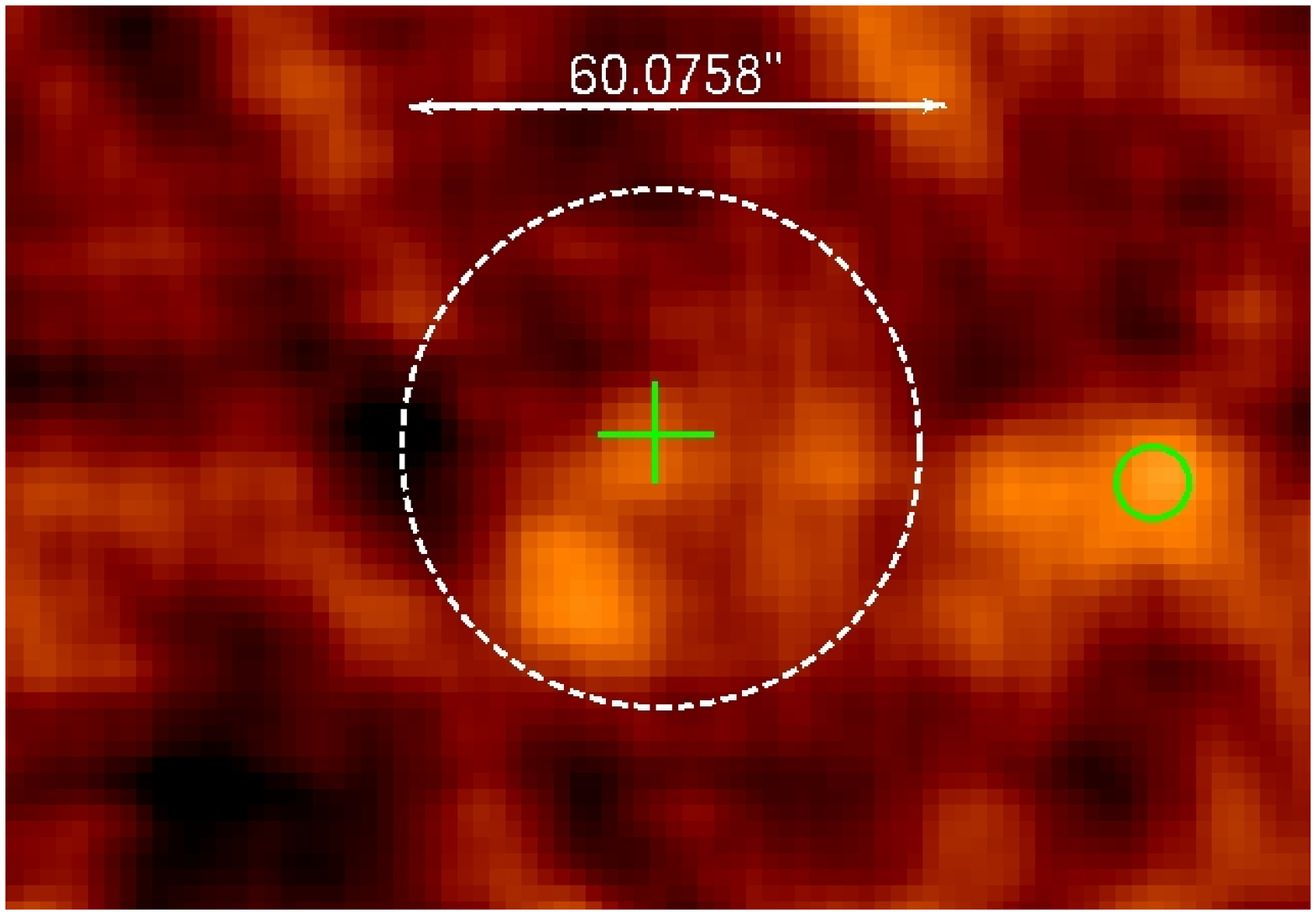}}
     \caption{FIRST cutouts around sources without FIRST counterparts. The
      dashed circle marks the NVSS source position and size, while the plus
      sign indicates the MegaZ optical galaxy location. The small circles mark
      the detected spots. Cases (a),(b) and (c) are examples of accepted matches
      and cases (d),(e) and (f) are examples of rejected ones. A boxcar sliding
      average filter has been applied to enhance the visibility.}
\end{figure*}

\subsection{Reliability, Completeness and Final Catalogue}
Table 1 lists the number of accepted radio matches for  different classes 
of radio sources found in the sky test area.
We provide results for both the real and the random samples,
as well as reliability estimates. Note that even though our sample is much
deeper and more than five times larger than the one of \citet{best05a}, we
have been able to compile a catalogue with  comparable reliability
($\sim$98.3\%).

Because the true number of radio sources is unknown, the completeness
is more difficult to estimate than reliability. Nevertheless, a
reasonable estimate of the completeneness can be obtained by
considering the MegaZ-LRGs that have an NVSS source within a distance
of 15 arcsec. These can be assumed to be a mixture of true matches,
random alignments from sources at different redshifts, and a
contribution from sources that are associated with other galaxies
which just happen, through clustering, to be close to the target
galaxy. By using the Monte-Carlo simulations in the sky test area to
estimate the latter two contributions, we can estimate the number of
true matches and hence the completeness.

In the sky test area there are 1765 MegaZ-LRGs with an NVSS source
within a distance of 15 arcsec. In comparison, there are an average of
225 sources within the same distance to the galaxies in
the random catalogues. Determining the contribution from clustered galaxies          
galaxies is more complicated. We attempted this in two ways.

First, we selected all NVSS-SDDS candidates with a single FIRST
component within 15 arsec, but which were rejected by the matching
algorithm. We then searched for other optical MegaZ galaxies in a 3
arcsec region around these FIRST radio positions (a standard search
radius for FIRST matching, e.g. \citealt{obric}).
From a total of 359 LRGs, 40 (11.1\%) had a nearby LRG with an
associated FIRST source. The corresponding fraction in the random
sample was 0\%. This means 11.1\% of the 1765
sources ($\sim$200) within 15 arcsec are due to clustering. 

To check this result, we estimated the excess number of  
LRG-LRG pairs over random  by integrating the angular correlation
function $w(\theta)$ given by \citet{ross} for the 2SLAQ catalog out
to 15 arcsec. This yielded value of 1.66, i.e.  $\sim 150$  extra sources
due to clustering. These two values are in rough agreement; the fact
that the first method gives a slightly larger number could be explained
by the fact that radio galaxies are more massive than the underlying LRG 
population, so we would expect them to be more strongly clustered.             
For our final completeness estimate,   we will assume that                        
180 sources are due to clustering.

The expected number of true matches is then estimated as
$1765-225-180=1360$. Our matching procedure detects 1309 objects, of
which 21.7 are predicted to be false detections. The completeness can
therefore be estimated as $100*(detected-false)/expected$ or
$100*(1309-21.7)/1360 \approx 95\%$.  This value is comparable to that
obtained by \citet{best05a}, for a much lower redshift sample.
We note that this completeness estimate applies only to the MegaZ-LRG
sample and does not take into account missing  radio-loud QSOs
or radio galaxies that do not meet the MegaZ-LRG colour cuts.
\footnote {A preliminary comparison with a complete spectroscopic catalogue
at similar redshifts indicates that $\sim$ 20\% of all galaxies with
$M_* > 10^{11} M_{\odot}$ are too blue to fall within the
MegaZ-LRG colour coundaries. Since radio AGN are biased towards the
higher mass galaxies, the corresponding percentage  should be smaller
for this population.}

Table 2 lists the final number of radio galaxies found in the MegaZ--LRG
sample and included in the catalogue. Our  catalogue is by far the largest
sample of radio galaxies of its kind. Out of a total of 14553 sources, the 
majority (78.6\%) have single counterparts in both FIRST and NVSS. A significant
fraction (11.5\%) are resolved by FIRST into multiple components. Only 253
(1.7\%) sources have multiple NVSS counterparts, to be compared with the
6.2\% found by \citet{best05a}. This is again expected, because the radio
galaxies lie at larger distances.

We note that  as for \citet{best05a}, a fraction of the
candidate sources (1093 objects, or 7.5\% of the sample) with multiple
NVSS or FIRST components were flagged as being too unreliable to classify
by automated means. These were classified visually by simultaneous
examination of SDSS, NVSS and FIRST images. Of these, 221 were accepted as
genuine sources. Finally, to ensure that the same NVSS source was not
associated with two different MegaZ--LRG galaxies, the complete catalogue
was checked. Such duplications arose for  560 candidates and all cases were visually
examined. Altogether, 564 positive matches (3.9\%) 
result from the visual classification process.

Finally, we note that it is important to distinguish between stars and galaxies
when compiling the radio galaxy catalogue, because
late M--type stars  have colours that are very similar to those of faint
LRGs. \citet{collist} used the same neural network technique used to derive
photometric redshifts,  to perform star/galaxy separation. The  neural net was 
given 15 different photometric parameters and trained using the 2SLAQ
spectroscopic sample. The MegaZ catalogue provides  a parameter,
$\delta_{\rm sg}$, which is the probability that a particular object is
a galaxy rather than a star. In the subsequent analysis, we only consider objects 
with $\delta_{\rm sg}>0.7$ (99.93\% of all the matched radio galaxies, 96.78\% of
all optical galaxies).

\subsection{Comparison Sample}
In the following sections, we will compare
the results obtained for the MegaZ--LRG catalogue of radio AGN
to those obtained for
galaxies selected from  the main spectroscopic galaxy sample, which have redshifts in 
the range $0.01<z<0.30$, with a mean redshift of $\langle z \rangle\sim0.14$. This
is an extension to the fourth data release of the original DR2
radio--matched catalogue of \citet{best05a}.

\begin{table}
\centering
\caption{Numbers of different classes of accepted MegaZ--LRG radio galaxies, 
as well as the corresponding numbers found in the random samples.
The numbers are for a  720 deg$^2$ patch of the total survey area.
Values in parentheses indicate the number of 
sources accepted using the spot--matching technique (see text).}
\begin{tabular}{@{}llrrr@{}}
\hline
NVSS type & FIRST type & MegaZ & Random & Reliab. \%\\
\hline
NVSS sing. & 0 comp.& 109(13) &  12.3(2.4) &  88.7\\
NVSS sing. & 1 comp.& 1068 & 7.8 &  99.3\\
NVSS sing. & 2 comp.& 88   & 1.2 &  98.7\\
NVSS sing. & 3 comp.& 38   & 0.3 &  99.2\\
NVSS sing. & 4$+$ comp.& 6 & 0.1 &  98.3\\
NVSS mult. &        & 24   & 0.8 & 96.7\\
\hline
Total      &       & 1333 & 22.5 & 98.3\\
\hline
\end{tabular}
\end{table}

\begin{table}
\centering
\caption{Total number of MegaZ--LRG  radio galaxies  
with 1.4 GHz flux above 3.5 mJy in the different classes.}
\begin{tabular}{@{}llrrr@{}}
\hline
NVSS type & FIRST type & MegaZ\\
\hline
NVSS sing. & 0  comp.& 1176\\
NVSS sing. & 1  comp.& 11445\\
NVSS sing. & $>$1 comp.& 1679\\
NVSS mult. &        & 253\\
\hline
Total      &       & 14453\\
\hline
\end{tabular}
\end{table}

\section{Radio--Loud AGN Evolution}
\subsection{Masses and $(K+e)$--corrections}
To derive stellar masses for each galaxy we used the {\it kcorrect} algorithm
(\citealt{blanton}), which finds the non--negative linear combination of spectral
templates that best matches the flux measurements of each galaxy in a $\chi^2$
sense. These templates are based on a set of \citet{bruzual} models and span a
wide range of star formation histories, metallicities and dust extinction, so
they can be used to estimate the mass--to--light ratio of a galaxy. This method    
yields stellar masses that differ by less than 0.1 dex on average from estimates
using other techniques, for example using the  4000{\AA} break strength and
$H\delta$ absorption index to estimate $M/L$ as proposed by \citealt{kauff03}.

In order to compare radio galaxies at different redshifts in a fair way, we need
to compute the volume over which the MegaZ sources are visible, given the
magnitude and colour cuts of the surveys as well as the 3.5 mJy radio flux limit.
In this work we generated $(K+e)$--corrections using a library of \citet{bruzual}
evolutionary stellar population synthesis models. LRGs constitute a nearly
homogeneous population of old, red, early--type galaxies with spectral
energy distributions that are reasonably well described by
a single stellar population (SSP) model that forms all its stars in an
instantaneous burst at high redshifts (we choose $z=9.84$) and then passively
evolves (\citealt{wake}). We use these models to compute $K$--corrections, which
account for  the redshifting of the spectra through the different rest--frame
bandpasses, and $e$--corrections, which account for the intrisic evolution in the
spectral energy distribution of galaxies because stars are younger at higher
redshifts. With these corrections in hand, we are able to predict the $u,g,r,i,z$
magnitudes as well as the radio luminosities of each MegaZ LRG as function of
redshift, which then allows us to determine the precise minimum ($z_{min}$) and
maximum ($z_{max}$) redshift that a particular object enters and then leaves
the radio--optical sample. The weight for each galaxy is then  $1/V_{max}$,
where $V_{max}$ is the integral of the comoving volume between $z_{min}$ and $z_{max}$.

As a check, we have computed the stellar mass function of the LRGs in the MegaZ
catalogue and we compare our results with stellar mass function of low redshift
massive galaxies drawn from DR4. This is shown in Figure 4. The  MegaZ--LRG
sample probes the high mass  end of the stellar mass function,
spanning a range in stellar mass from  $10^{11}$ to $ 10^{12.1}M_{\odot}$. Over
this range in mass, there is essentially no evolution in the stellar mass
function out to $z \sim 0.55$. This is in accord with recent results presented
by \citet{brown07} and \citet{cool}, who show that the upper end of the
stellar mass function has apparently changed very little out to a redshift of
$z \sim 0.9$.

\begin{figure}
\includegraphics[width=84mm]{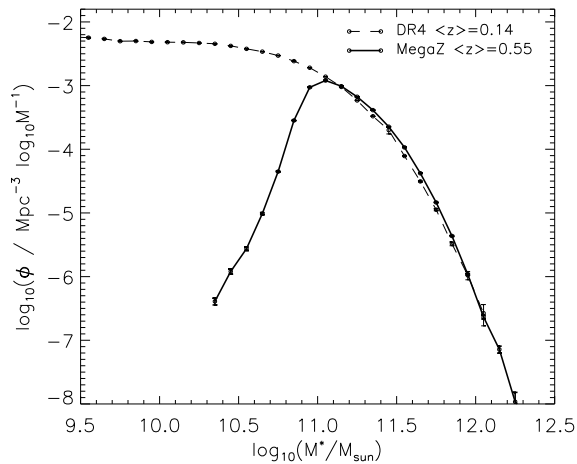}
\caption{Mass function for galaxies in the MegaZ--LRG (solid thick line) and
the DR4 sample (dashed thin line).}
\end{figure}

\subsection{Radio Luminosity Function}
A standard technique for quantifying the rate of evolution of a population of
galaxies is to compare the radio luminosity function of these objects at two
different epochs. In this section, we will determine the evolution of the
radio--loud AGN population by comparing the  luminosity function of MegaZ--LRG
radio galaxies with that  derived from  the SDSS DR4 sample.

These functions were calculated for radio sources above a 1.4 GHz flux density
limit of 3.5 mJy using the standard  $1/V_{max}$ formalism (\citealt{schmidt}).
We estimate $V_{max}$ for each galaxy as the maximum comoving volume within
which it would be visible given the radio flux and optical magnitude
limits and the colour cuts used to define the  MegaZ--LRG sample (see Section 2.1).
We have included both $K$--corrections and $e$--volutionary corrections in this
computation. These corrections were described in the previous section. Radio
luminosities are calculated using the formula
\begin{equation}
\rm log_{10}[P_{1.4 GHz}]=log_{10}[4\pi D^{2}_{L}(z)S_{1.4 GHz}(1+z)^{\alpha -1}]
\end{equation}
where $\rm D_{L}$ is the luminosity distance in the adopted cosmology,
$\rm S_{1.4 GHz}$ is the measured radio flux, $\rm (1+z)^{\alpha -1}$ is the
standard $k$--correction used in radio astronomy, and $\alpha$ is the radio
spectral index ($S_\nu \propto \nu^{-\alpha}$) for which we adopted a value of
0.7, as is usually assumed for radio galaxies (\citealt{condon02}).

The normalization of the luminosity function requires knowledge of the precise
intersection area of all three surveys: NVSS, FIRST and the SDSS DR4 photometric
sample. We used  the footprint services of the
Virtual Observatory\footnote{http://www.voservices.net/footprint}
(\citealt{budav}) to derive a total area of 6164.6 deg$^2$. A further
8.3 deg$^2$ was subtracted to avoid noisy regions around bright NVSS sources,
resulting in an effective area of 6156.3 deg$^2$. In the MegaZ sample, the
maximum variation of the mean redshift as we go from our second luminosity bin
to the highest luminosity one, is $\sim0.025$. If we include the first bin, this
difference increases to a value $\sim0.08$. We nevertheless decide to keep the
lowest luminosity bin to avoid introducing further luminosity cuts.

In Figure 5 we plot the radio luminosity function measured for the MegaZ--LRG
catalogue and the results are also tabulated in Table 3. For comparison we
reproduce the luminosity function calculated by \citet{sadler07} for the 2SLAQ
LRG sample. There is very good agreement, but our new determination
has considerably smaller error bars, because our sample is very much larger.
This is  especially important for determining the number density of radio
sources with radio powers greater than
log$_{10}$(P$_{\rm 1.4 GHz}$)$=26$ W Hz$^{-1}$. Also plotted in the figure are two
determinations of the radio luminosity function from SDSS DR4: the first without
color restrictions, and the second restricted to red galaxies as defined by a
luminosity dependent $g-r$ color cut found by fitting bi-Gaussian functions to
the distribution of $g-r$. The details of the method are explained by \citet{li}.
For reference purposes, we also show the RLF obtained by \citet{best05a} for the
DR2 sample.

The evolution of the luminosity function of a population of objects is often
characterized in terms of simplified density or luminosity evolution models. In
a pure density evolution model, the underlying luminosity distribution of the
radio sources  remains fixed, but the number density of the sources increases
at higher redshifts. This might be the case if a larger fraction
of galaxies harbour radio jets at higher redshifts, but the intrinsic properties
of the jets do not evolve with time. In a pure luminosity evolution model, there
is a fixed population of radio AGN with luminosities that dim as the radio
sources age.

In the two right--hand panels of Figure 5, we plot the ratios $f_{\rm PDENS}$
and $f_{\rm PLUM}$ as a function of radio luminosity. $f_{\rm PDENS}$ is
the ratio of the comoving number densities of
radio sources in the MegaZ and DR4 catalogues at fixed radio luminosity, while
$f_{\rm PLUM}$ is the factor by which the low redshift number densities should
grow in order to match the values found at higher redshift. If either model is a
correct description of the data, one should find constant $f$ values as a
function of radio luminosity. We find that neither model is a good description
of the data. Both plots show a sharp increase near
log$_{10}$(P$_{\rm 1.4 GHz}$)$\sim25$ W Hz$^{-1}$.

In Figure 5, we also plot the radio luminosity function for steep--spectrum
sources at z=0.55 predicted by the luminosity/density evolution model of
\citet{dunpeack}. This model is based on high frequency data, and has been shown
to provide an accurate description of the radio source counts and redshift
distributions. At the redshifts corresponding to our MegaZ-LRG sample,  the models were only strongly
constrained above $10^{26}$ W Hz$^{-1}$ due to the high flux limits of their radio
samples, and were then extrapolated to the lower radio powers. We note that
\citet{dunpeack} provide a range of different parametrizations in their paper, but
all give very similar results at moderate redshifts. As can be seen, our data
matches the \citet{dunpeack} model reasonably well. In partcular, the
extrapolation to lower radio power appears to provide a reasonable fit to our data,
but our estimate appears to be 0.2-0.3 dex lower that that of \citet{dunpeack}
over the luminosity range from $10^{25}$ to $10^{26.5}$ W Hz$^{-1}$.
\begin{figure*}
\includegraphics[width=150mm]{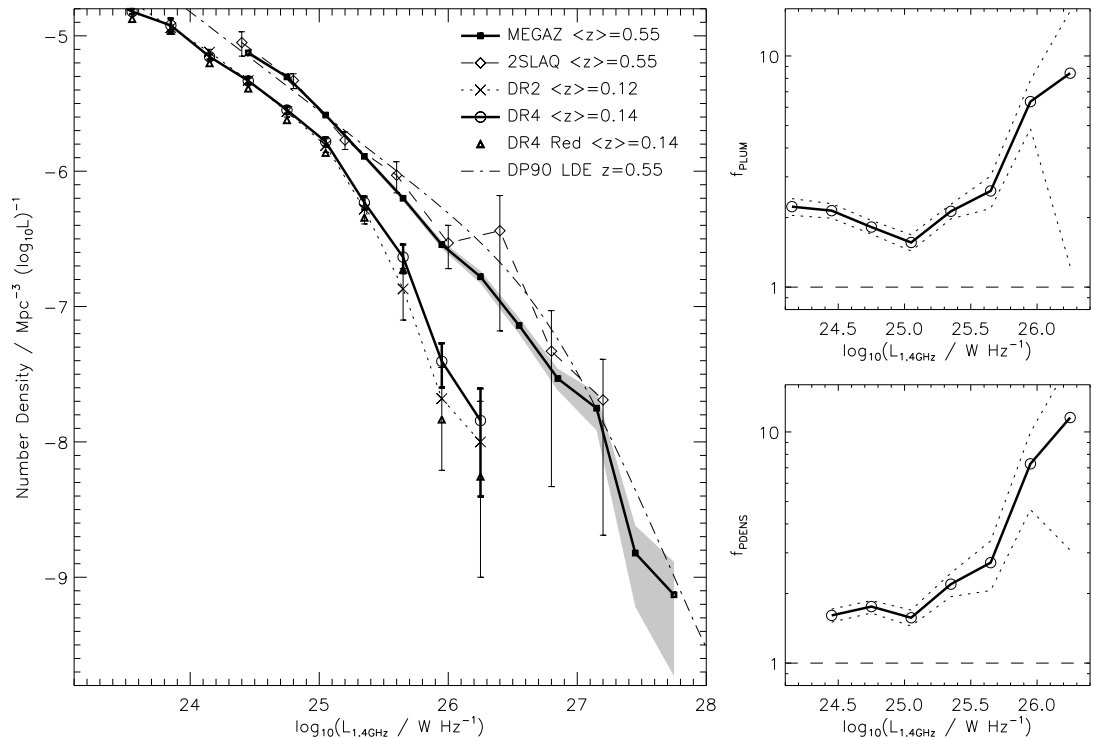}
\caption{The radio luminosity function at 1.4 GHz derived for MegaZ--LRG and DR4
samples (thick line with square and circle symbols, respectively). An additional
DR4 RLF restricted only to red sequence galaxies (triangles) is also plotted.
Likewise shown are the calculations by \citet{sadler07} for 2SLAQ (thin line and diamond
symbols), and by \citet{best05a} for SDSS DR2 (dashed line and crosses). Results
are in good agreement at both redshift intervals. The luminosity function based
on the luminosity/density evolution model of \citet{dunpeack} is also
overplotted (dash--dotted line). The right panels show the factors by which
luminosity and density should increase in order to match the DR4 curve at
$\langle z \rangle\sim0.14$ with the MegaZ--LRG curve at
$\langle z \rangle\sim0.55$. No evolution scenarios are indicated by the
horizontal dashed lines. Error bars are poissonian, except for the right upper
panel where they are calculated via statistical bootstrapping of each sample.}
\end{figure*}

It is important to note that if radio galaxies fall outside the MegaZ--LRG colour
selection criteria, they will be missed in our radio luminosity function determination.
Such an effect is visible for our low redshift luminosity function at radio
powers above $10^{26}$ W Hz$^{-1}$; the luminosity function of bright sources
in the red galaxy population is 0.1-0.2 dex lower than for the DR4 sample as a whole.
Neither the DR4 sample nor the MegaZ-LRG sample includes radio--loud quasars, which would
likely increase the amplitude of the radio luminosity functions by $\sim 0.1$ dex above
a radio luminosity of $10^{25}$ W/Hz$^{-1}$. Nevertheless, it is clear from our comparisons  
that  a large fraction (i.e $>50$\%)  of all radio galaxies at $z=0.55$ do reside
in the LRG population. In the rest of the paper, we will be comparing radio AGN
{\em fractions} at low and high redshifts, and we will show that such fractional
estimates are much less sensitive to any incompleteness in our sample.

\begin{table}
\centering
\caption{The MegaZ--LRG luminosity function for radio--loud AGN at 1.4 GHz}
\begin{tabular}{@{}llrrr@{}}
\hline
log$_{10}$(P$_{\rm 1.4 GHz}$) & log$_{10}$($\rho$) & N\\
(W Hz$^{-1}$) & (Mpc$^{-3}$ (log$_{10}$L)$^{-1}$) \\
\hline
24.45  & -5.13$^{+0.02}_{-0.02}$ & 2218 \\
24.75  & -5.30$^{+0.01}_{-0.01}$ & 5438 \\
25.05  & -5.59$^{+0.01}_{-0.01}$ & 3506 \\
25.35  & -5.89$^{+0.02}_{-0.02}$ & 1742 \\
25.65  & -6.20$^{+0.02}_{-0.02}$ &  840 \\
25.95  & -6.54$^{+0.03}_{-0.03}$ &  365 \\
26.25  & -6.78$^{+0.04}_{-0.05}$ &  185 \\
26.55  & -7.14$^{+0.06}_{-0.06}$ &   87 \\
26.85  & -7.53$^{+0.07}_{-0.09}$ &   41 \\
27.15  & -7.75$^{+0.12}_{-0.16}$ &   18 \\
27.45  & -8.82$^{+0.20}_{-0.40}$ &    3 \\
27.75  & -9.13$^{+0.24}_{-0.60}$ &    1 \\
\hline
\end{tabular}
\end{table}

\subsection{Fraction of Radio--Loud AGN}
We are now in a position to investigate the relation between radio--loud AGN and
their host galaxies. Over the redshift range covered by the MegaZ--LRG sample
($0.4 < z < 0.8$), radio sources have 1.4 GHz powers well above
$10^{24}$ W Hz$^{-1}$ (the faintest radio galaxies in our sample are
$\sim10^{24.4}$ W Hz$^{-1}$). It is well known that radio emission above this
luminosity is rarely associated with star-formation, but rather an AGN
(\citealt{sadler02}; \citealt{best05a}). In addition, luminous red galaxies are
photometrically defined so that the majority of star--forming galaxies are
excluded, so we can safely assume that our sample is dominated by an AGN population.

\begin{figure*}
\centering
\includegraphics[width=160mm]{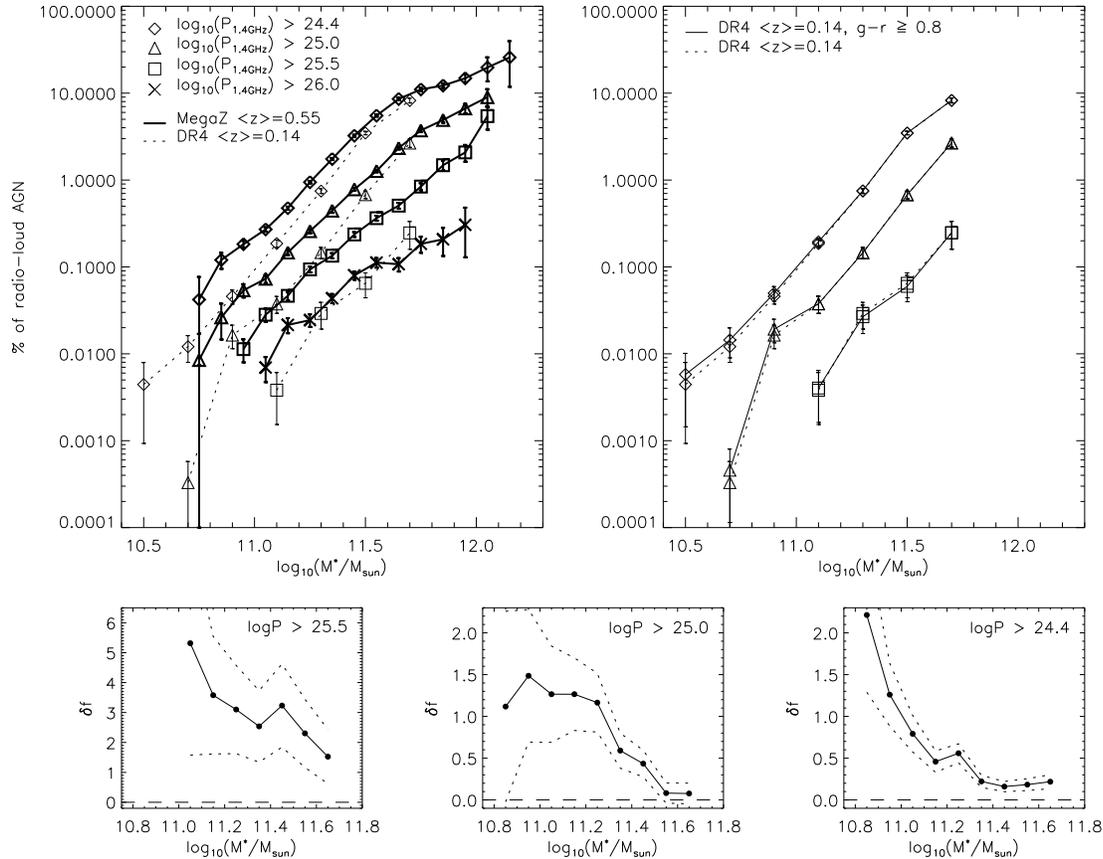}
\caption{The left upper panel shows the percentage of galaxies have radio luminosities
above a given threshhold, as a function of stellar mass. Results are shown for the 
high redshift MegaZ--LRG sample (thick solid lines), and for  the low redshift
SDSS DR4 sample (thin dashed lines). The right upper panel shows the 
radio--loud fractions for the SDSS DR4 sample, but including a colour constraint of
$g-r\geq0.8$. Lower panels show the relative variation $\delta f$ between high
redshift (MegaZ) and low redshift (DR4) radio--loud fractions as a function of
stellar mass. Errors are poissonian and are propagated via standard error
propagation theory.}
\end{figure*}

Figure 6 shows the fraction of MegaZ LRGs with radio luminosities above
different threshholds, as a function of their stellar mass. The fraction of LRGs
with radio AGN more luminous than $10^{24.4}$ W Hz$^{-1}$ rises from $\sim$0.04\% for
galaxies with stellar masses around $10^{10.8} M_{\odot}$, to $\sim$25\% for the
most massive LRGs with $ M_* \sim 10^{12.2} M_{\odot}$. Results for radio
power cuts of $10^{25}$ W Hz$^{-1}$, $10^{25.5}$ W Hz$^{-1}$ and
$10^{26}$ W Hz$^{-1}$ are also included in the plot and a similar mass dependence
is found. The dotted lines connecting light symbols on the plot show the
corresponding dependence  of the radio loud fraction on stellar mass for the
low--redshift SDSS DR4 sample. We find that there is an increase at higher
redshifts in the fraction of of radio--loud AGN hosted by galaxies of all stellar
masses. The increase is larger at higher radio luminosities and for galaxies
with lower stellar masses. This is shown in more detail in the bottom three
panels of Figure 6, where we plot the relative ratio $\delta f$ between the high
redshift and low redshift data ($\delta f=(f_{\rm MegaZ}-f_{\rm DR4})/f_{\rm DR4}$)
as a function of mass and for our three radio luminosity thresholds.

Once again it is important to recall that the DR4 sample includes all galaxies
regardless of colour, whereas the MegaZ sample is restricted to red galaxies.
\citet{best05b} showed that the  probability for a galaxy to host a
radio--loud AGN was not influenced by its stellar population. In the top right
panel of Figure 6 we compare the radio--loud AGN fraction versus mass relation
for the full DR4 sample to the results obtained if the sample is restricted to
galaxies with $g-r$ colours greater than 0.8. As can be seen, the results are
almost identical. The effect of the colour cut is much smaller than the
evolutionary trends we find between the low redshift and MegaZ LRG sample.
\begin{figure*}
\centering
\includegraphics{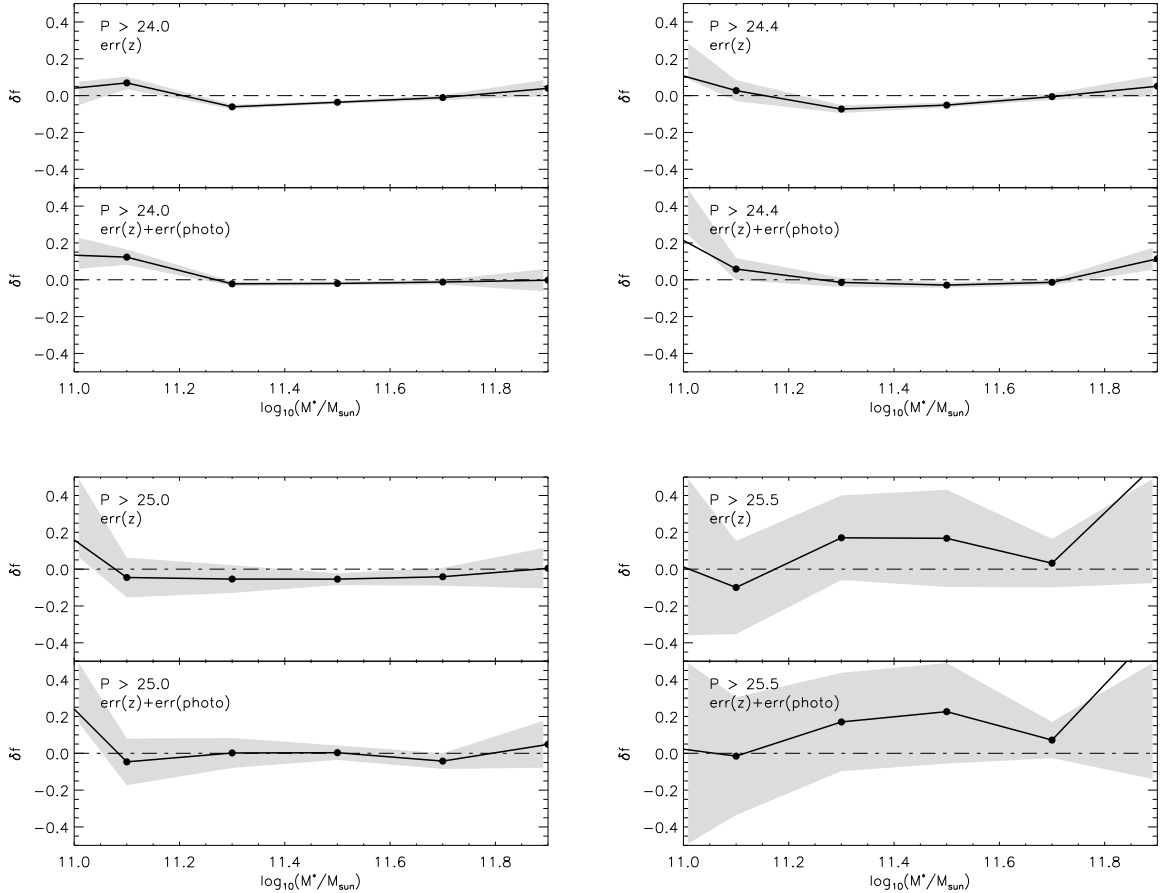}
\caption{The difference between the true radio--loud AGN fraction 
and the one calculated for the artifically redshifted samples is plotted
as function of stellar mass. Results are shown for increasing luminosity cuts.
The solid line shows the mean difference for the 25 random samples, and the
gray area is the 1-$\sigma$ scatter in this variation. Very few sources have masses
higher than $10^{11.7} M_{\odot}$.}
\end{figure*}

To check whether the increase in the fraction of radio--loud AGN with mass is not
simply the result of the fact that more massive galaxies harbour more massive
black holes and hence produce more radio emission, we have also calculated
the fraction of radio--loud AGN that are above a given limit in radio luminosity
per unit stellar mass. For our sample, this is equivalent to choosing only those
sources that produce radio emission above a fixed fraction of the Eddington
limit. For reasonable values of the chosen limit (e.g.
$\rm L/M^*>10^{13.3} W Hz^{-1} M^{-1}_{\odot}$), the strong mass dependences
persists, although the slope of the relation is somewhat shallower.

\subsection{Other systematic effects}
Photometric redshift codes have improved vastly over the years and the resulting
redshift estimates are useful for a broad range of studies. Even though the
neural network photometric redshift estimator used in the MegaZ catalogue has been
throughfully calibrated using  spectroscopic redshifts, it is important to test
the senstivity of our results to photo--z errors.
Such errors could translate into systematic offsets in our estimates of
stellar mass. In addition, the errors in the Sloan $u,g,r,i,z$ magnitudes 
increase with redshift, so one might worry that this would translate into apparent
(as opposed to real) evolution of the radio loud fraction with redshift.

In order to test for such effects, we artificially redshifted our SDSS DR4
sample from their observed redshifts near $\langle z \rangle\sim0.14$ out to the
mean redshift of the MegaZ--LRG sample, $z=0.55$. To do this, we applied the
{\it kcorrect} algorithm in conjunction with the $e$--corrections described above.
We then added an error to the redshift 
using the distribution of $z_{phot}-z_{spec}$ of 2SLAQ galaxies. In practice, for
each MegaZ LRG, we extracted the 100 closest neighbors in $g,r,i,z$ magnitude
space in the 2SLAQ catalogue and we draw the redshift error, $z_{phot}-z_{spec}$,
at random from this set of galaxies. Magnitude errors were assigned to
each DR4 object in the same way. We then ran our code  for estimating
stellar mass on this ``artificially redshifted'' data set, first applying just the 
redshift error, and second, incorporating errors on both the photometry and the
redshifts. The whole procedure was repeated 25 times.

The solid line in Figure 7 shows the difference between the mean fraction of
radio--loud AGN measured in the 25 artifically redshifted samples and the ``true''
value for the DR4 galaxies as a function of stellar mass. The difference $\delta f$
is expressed in fractional terms, i.e. $\delta f = 
(\bar{F}_{\rm artificial} -F_{\rm true})/ F_{\rm true}$.
The shaded contours respesent the 1$\sigma$ scatter among the 25 samples. It can
be seen that small systematic shifts in the derived radio loud fractions do
occur at the two ends of the stellar mass distributions.
They also occur in our estimates
of the fraction of the most luminous radio galaxies. On the whole, these effects
are small ($< 20$\%).  Note also that $10^{25.5}$ W Hz$^{-1}$ sources are much
rarer in the DR4 catalogue than in the MegaZ sample, so the bottom right panel
overestimates the size of the true effect at this radio luminosity and is likely
to apply only to  very much higher luminosity cuts at $z=0.55$.
\begin{figure*}
\centering
\includegraphics{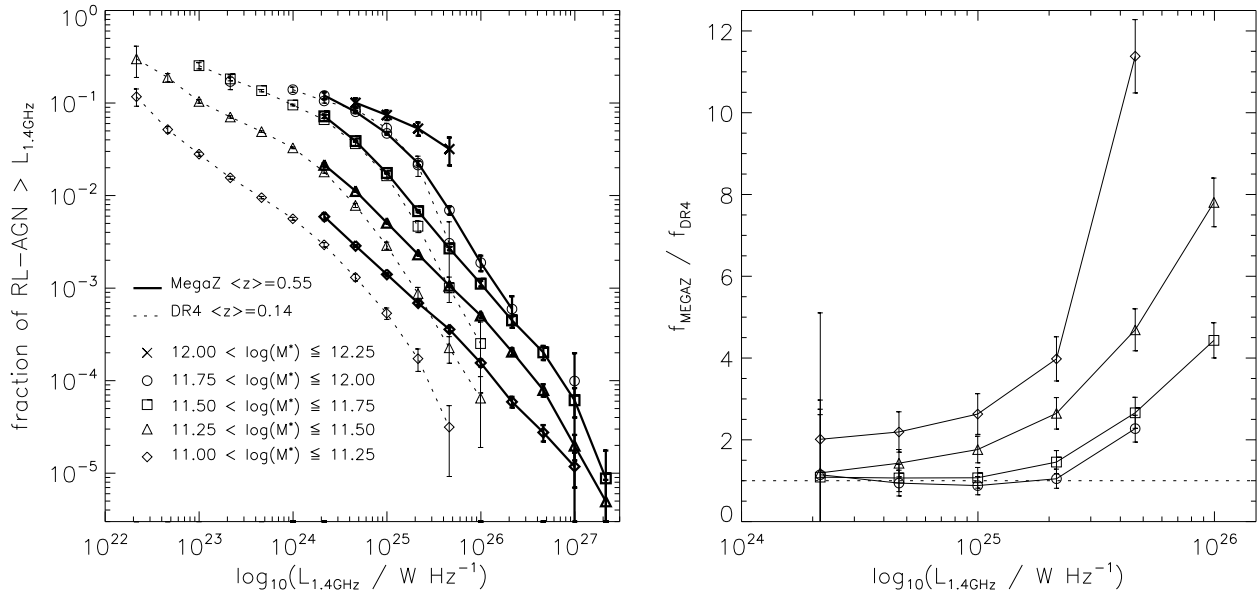}
\caption{Integral bivariate radio luminosity--mass function in several bins of
stellar mass (symbols) for high redshift MegaZ--LRG sources (thick solid lines) and low
redshift SDSS DR4 sources (thin dashed lines). The bivariate function gives the
fraction of radio--loud AGN brighter than a given radio luminosity, as a function
of the luminosity. The right panel shows the ratios of the  bivariate functions
derived for the MegaZ--LRG sample to those derived for the SDSS DR4 sample. 
These results show that radio sources evolve strongly across cosmic time
depending on their stellar mass and radio luminosity. The dashed horizontal line
marks the no--evolution scenario.}
\end{figure*}

\subsection{Bivariate radio luminosity--mass function}
The first determination of the bivariate radio--optical luminosity function was
carried out by \citet{auriem} for a sample of 145 galaxies of type E and S0, and
there were subsequent studies by \citet{sadler89} and \citet{ledlowowen}.This
(integrated) function measures the probability that a galaxy in a given optical
absolute magnitude range hosts an AGN  with radio power above a certain value. 
\citet{best05b} calculated the bivariate radio luminosity--stellar mass function
of radio sources in the SDSS DR2, and parameterized the behaviour using a broken
power with characteristic luminosity $P_{\star}=2.5\times10^{24}$W Hz$^{-1}$. A
single fitting function was found to hold for for all masses below
$10^{11.5}M_{\odot}$.

Our MegaZ radio galaxy sample is over 100 times larger than the one of \citet{auriem}
and picks up close to the break luminosity found by \citet{best05b}. Following
\citet{best05b}, we will use stellar mass instead of optical luminosity.  We
can thus interpret our calculation as the fraction of radio--loud AGN of a given mass
emitting above a certain certain radio power. Figure 8 shows the bivariate
functions at high and low redshift in mass bins from $10^{11} M_{\odot}$ to
$10^{12.25} M_{\odot}$. The most striking result shown in this plot is that for
galaxies of all stellar masses, the high redshift fractions lie only slightly
above the low redshift ones at faint radio luminosities, but there has been much
stronger evolution at higher radio luminosities.
This is shown in more detail in the right hand panel of Figure 8, where we plot
the ratio of the radio--loud AGN fraction derived for the MegaZ LRGs to that
derived for the DR4 galaxies, as a function of radio power. We show results for
the four mass bins for which there is reasonable overlap between the two surveys
(there many fewer galaxies with $12<\log M_*<12.25$ in the DR4 sample, because
the volume covered is much smaller). It is striking that for all the mass bins,
the ratio remains approximately constant up to a radio luminosity of
$10^{25}$ W Hz$^{-1}$ and then turns up sharply. As discussed in Section 1,
$10^{25}$ W Hz$^{-1}$ corresponds to the approximate dividing line between
low excitation and high excitation sources, and also delineates the boundary
between strongly--evolving and less strongly--evolving radio sources found in
previous studies. The main new result shown in this plot is that the strongest
evolution in the radio source population apperently occurs
{\em in the lowest mass galaxies}.

\begin{figure}
\includegraphics[width=84mm]{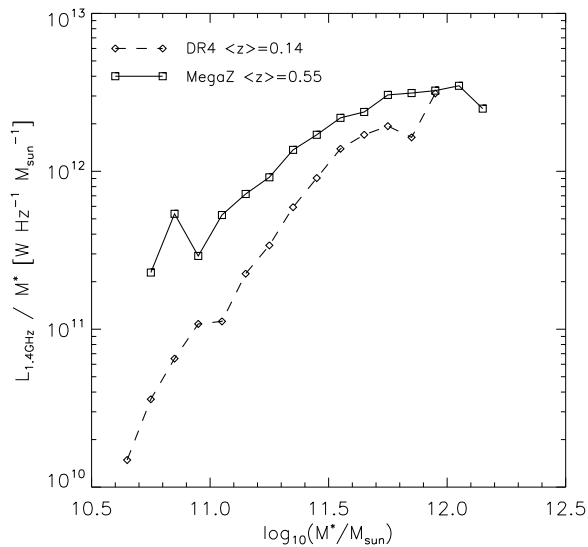}
\caption{Integrated volume--weighted radio luminosity per 
unit stellar mass, as function of stellar mass, for MegaZ--LRG
radio--loud AGN (solid line with square symbols) and DR4 sources (dashed line
with diamond symbols).}
\end{figure}

Another way of showing the same thing is to plot the integrated radio emission
per unit stellar mass for galaxies of different stellar masses in the DR4 and
MegaZ samples. \citet{heckman} used the volume--weighted distribution of
[OIII] 5007 line luminosity per unit black hole mass to study the average rate
at which black holes are accreting and growing. They find that low mass black
holes are currently growing at substantially higher rates than high mass black
holes. \citet{best05b} extended this analysis by computing the integrated radio
luminosity per unit black hole mass, demonstrating that radio and emission line
luminosities are produced in black holes of quite different masses (see Figure
11 in \citealt{best05b}).

In the present work, we do not have direct estimates of black hole mass, so we
carry out the computation as a function of stellar mass. In Figure 9 we plot the
volume--weighted radio luminosity per unit stellar mass, as a function of $\log M_*$,
for our high and low redshift samples. The plot is constructed by taking at each
mass bin, the ratio of the integrated (and weighted) radio luminosity, to the
integrated (and weighted) stellar mass in that bin. As can be seen, radio
activity has been boosted at all masses at high redshift, but the boost factor
is considerably higher for low mass galaxies.

\section {The stellar masses and colours of the host galaxies of radio--loud AGN}
It has long been known that powerful high redshift radio galaxies are hosted
by very massive galaxies with predominantly old stellar populations. This
conclusion arises from the fact that there is a well defined relation between
the K--band magnitude of radio galaxies and redshift (the famous K--z relation)
that agrees with the predicions of a passively evolving model galaxy of high mass
($>10^{11} M_{\odot}$), and that there is relatively little scatter about this
relation for the most luminous radio sources (e.g. \citealt{lilly}; \citealt{jarvis}).

So far we have phrased our analysis of the relationship between radio AGN and
their host galaxies in terms of the {\em probability} for a galaxy of
a given mass to host such an AGN. We have shown that this {\em probability}
evolves most strongly in low mass galaxies. Note, however, that the fraction of
radio loud AGN is a strongly increasing function of galaxy mass both at
$z=0.55$ and at $z=0.14$. This means that radio galaxies in massive elliptical
hosts always dominate {\em by number}.
\begin{figure}
\includegraphics[width=84mm]{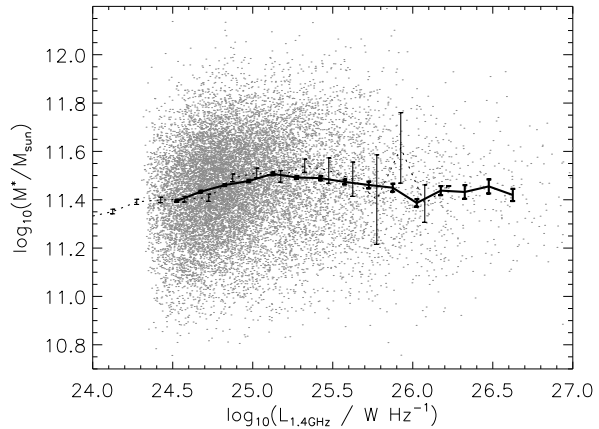}
\caption{Median stellar mass of radio--loud AGN in the MegaZ--LRG (solid line) and
SDSS DR4 (dotted line) samples as function of radio power. Gray dots represent the
actual values for each MegaZ galaxy and errors are calculated by the bootstraping
technique.}
\end{figure}

This is shown in more detail in Figure 10 where we plot the median stellar
masses of both the MegaZ and the SDSS DR4 radio AGN as a function radio
luminosity. As can be seen, radio AGN are hosted by galaxies with median masses
of $\sim 3 \times 10^{11} M_{\odot}$ in both surveys. There is a weak increase
in the median mass at higher radio luminosities.

In Figure 11, we plot the colour distributions of the MegaZ radio AGN. One must
of course bear in mind that the MegaZ sample is selected so as to occupy a
restricted range in colour space, but we can nevertheless check whether there
are any colour differences between different types of radio AGN. We split the
sample into 3 different redshift bins and show results for 
low luminosity ($< 10^{25.3}$ W Hz$^{-1}$) and high luminosity
($> 10^{25.3}$ W Hz$^{-1}$) radio sources. As was found previously by \citet{smith},
there is a tendency for the more powerful radio galaxies to have slightly bluer
colours at fixed redshift, but the effect is very weak.

In summary, we conclude that the majority of both low luminosity and high
luminosity radio AGN are hosted by massive (3 $L_*$) galaxies with red colours.
This is in good agreement with past work on this subject.
\begin{figure}
\includegraphics[width=84mm]{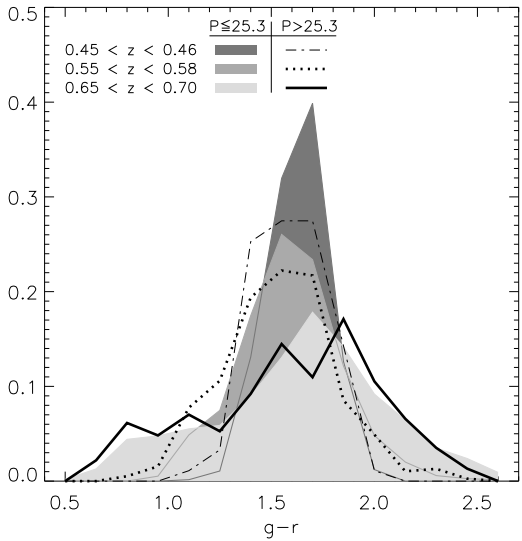}
\caption{Colour distribution of MegaZ--LRG in three different redshift bins, split
into a high luminosity population of radio sources with
log$_{10}$(P$_{\rm 1.4 GHz}$)$\leq25.3$ W Hz$^{-1}$, and a low luminosity
population with log$_{10}$(P$_{\rm 1.4 GHz}$)$>25.3$ W Hz$^{-1}$.}
\end{figure}

\section{Summary and Discussion}
The main results of the present work can be summarized as follows:\\

A catalogue of 14453 radio--loud AGN with 1.4 GHz fluxes above 3.5 mJy in the
redshift range $0.4<z<0.8$, has been constructed from the cross--correlation of
NVSS and FIRST radio source catalogues with the MegaZ--LRG catalogue of luminous
red galaxies. The vast majority of the radio AGN are single component sources in
both NVSS and FIRST. However, there is a signifiant fraction of objects without
high S/N FIRST detections, which are required for accurate identification 
of the optical counterpart. We thus introduced a method for analyzing the FIRST
radio maps around the candidate positions of these sources. This allowed us to
dig deeper into the FIRST survey and use lower S/N detections to pinpoint the
location of the host galaxy.

We have presented a new determination of the luminosity function of radio--loud
AGN at $z \sim 0.55$ that is in excellent agreement with other results in the
literature, but with notably smaller error bars. By comparing our radio
luminosity function at $z \sim 0.55$ to that derived for a large sample of
nearby radio AGN from the SDSS DR4, we find compelling evidence for strong cosmic
evolution of radio sources. The comoving number density of radio AGN with
luminosities less than $10^{25}$ W Hz$^{-1}$ increases by a factor of $\sim 1.5$
between $z=0.14$ and $z=0.55$. At higher lumiosities, this factor increases
sharply, reaching values $\sim 10$ at a radio luminosity of $10^{26}$ W Hz$^{-1}$.
Neither a pure luminosity evolution nor a pure density evolution scenario
provides a good description of the data.

We then turn to an analysis of how the relation between radio AGN and their host
galaxies evolves with redshift. The fraction of galaxies with radio luminosities
above a given threshhold is a steeply increasing function of stellar mass at
both $z \sim 0.1$ and at $z \sim 0.55$. The fraction of radio loud AGN increases
with redshift and this increase is largest at the highest radio luminosities and
also for lower mass galaxies. We have also calculated the bivariate radio
luminosity--mass function at $z \sim 0.14$ and $z \sim 0.55$. Its shape does
not appear to depend on mass at either redshift, but there is strong evolution
in the shape at the {\em bright--end} of the function. The fraction of galaxies brighter
than  $10^{25}$ W hz$^{-1}$ declines with significantly shallower slope at higher
redshifts. Within the range $10^{25-27}$ W hz$^{-1}$, simple power law fits to
the DR4 bivariate function in Figure 8 ($\rm frac \propto \rm L_{\rm 1.4 GHz}^{\gamma}$),
produce slopes $\gamma$ of $-1.84\pm0.21$, $-1.66\pm0.02$ and $-1.83\pm0.04$ for the
first, second and third mass bins. In MegaZ, we obtain
$-1.05\pm0.02$, $-1.16\pm0.05$ and $-1.20\pm0.02$ for the same luminosity range
and mass bins.

In short, two main conclusions have emerged from our analysis:
\begin{itemize}
\item There is a characteristic luminosity  of $\sim 10^{25}$ W Hz$^{-1}$
below which the radio source population appears to evolve only very weakly with
redshift. Above this characteristic luminosity, there is strong evolution, with
the most powerful radio sources undergoing the largest increase in co--moving
number density. These results are in broad agreement with past studies.
\item The strongest evolution in the fraction of galaxies that host radio--loud
AGN takes place in the lower mass galaxies in our sample.
\end{itemize}

The most plausible explanation of these trends is that there are two classes of
radio galaxy, likely associated with the high excitation/low excitation ``dichotomy")
that have different fuelling/triggering mechanisms and hence evolve in different ways.
This has also been argued by \citet{tasse}, whose analysis of a sample 1\% of the
size of ours in the XMM-LSS region provided hints of similar evolutionary properties.
As discussed before, it has been hypothesized that the class of low luminosity AGN is
likely associated with massive ($ \sim 10^{13} M_{\odot}$) dark matter halos
with quasistatic hot gas atmospheres. The abundance of such halos is predicted
to remain approximately constant out to redshifts $\sim 1$ (\citealt{mowhite})
and this is broadly consistent with the very weak evolution we see in this
population. These ``hot--halo" triggered sources at $z\sim0.5$ would then
produce radio emission that roughly compensates the radiative losses of the hot
gas.

The reason for the remarkably strong evolution in the comoving abundance of
high--luminosity radio AGN is somewhat more difficult to understand. As shown
in Figure 8, this evolution is most dramatic in galaxies at the low stellar mass
end of our sample. It is thus tempting to link this population with the strongly
accreting (and evolving) population of luminous quasars, whose space densities
also increase by factors of more than 10 over the redshift range studied in this
paper. Recent work has shown that quasars are hosted in dark matter halos of
masses $\sim 10^{12} M_{\odot}$, independent of redshift or the optical
luminosity of the system (e.g. \citealt{croom}; \citealt{shen}). Only $\sim 10$\%
of optical quasars are radio--loud, however, and this raises the question as to
whether there is simply a short radio--loud phase during the lifetime of every
optical quasar, or whether optically luminous AGN and powerful radio galaxies
are triggered under different physical conditions. If, as in quasars, these
sources are not regulated by cooling from a surrounding hot gaseous halo, but by
other mechanisms like interactions/mergers, then we would not expect to find
them in an equilibrum state like in low excitation AGNs.

One way to shed further light on these matters is to compare the clustering
properties of quasars to those of the luminous radio galaxies in our sample. If
their clustering properties are identical, this would favour the hypothesis that
all quasars experience a radio--loud phase. Recently, \citet{kbh} found that nearby
radio-loud AGN were more strongly clustered than matched samples of  radio-quiet AGN
with the same black hole masses and extinction corrected [OIII] line luminosities.
It will be interesting to see if the same conclusion holds at higher redshifts and
for more powerful systems. This will be the subject of a future paper.

\section*{Acknowledgments}
We would like to thank the Max Planck Society for the financial support provided
through its Max Planck Research School on Astrophysics PhD program. We also
thank T. Heckman, C. Li, V. Wild, and N. Drory  for valuable support and suggestions.
The research uses the SDSS Archive, funded by the Alfred P. Sloan Foundation,
the Participating Institutions, the National Aeronautics and Space Administration,
the National Science Foundation, the US Department of Energy, the Japanese
Monbukagakusho and the Max Planck Society. The research project uses the NVSS
and FIRST radio surveys, carried out using the National Radio Astronomy Observatory
Very Large Array. NRAO is operated by Associated Universities Inc., under cooperative
agreement with the National Science Foundation.

\label{lastpage}
\end{document}